\documentclass[10pt, twocolumn, pre, aps, superscriptaddress, showpacs]{revtex4-1}
\usepackage{amsmath, graphicx, subfigure, tikz}
\begin{document}
    \title{The Chiral Potts Spin Glass in d=2 and 3 Dimensions}
    \author{Tolga \c{C}a\u{g}lar}
    \affiliation{Faculty of Engineering and Natural Sciences, Sabanc\i\ University, Tuzla, Istanbul 34956, Turkey}
    \author{A. Nihat Berker}
    \affiliation{Faculty of Engineering and Natural Sciences, Sabanc\i\ University, Tuzla, Istanbul 34956, Turkey}
    \affiliation{Department of Physics, Massachusetts Institute of Technology, Cambridge, Massachusetts 02139, USA}
    \pacs{75.10.Nr, 05.10.Cc, 64.60.De, 75.50.Lk}



    \begin{abstract}
The chiral spin-glass Potts system with $q=3$ states is studied in
$d=2$ and 3 spatial dimensions by renormalization-group theory and
the global phase diagrams are calculated in temperature, chirality
concentration $p$, and chirality-breaking concentration $c$, with
determination of phase chaos and phase-boundary chaos.  In $d = 3$,
the system has ferromagnetic, left-chiral, right-chiral, chiral
spin-glass, and disordered phases.  The phase boundaries to the
ferromagnetic, left- and right-chiral phases show, differently, an
unusual, fibrous patchwork (microreentrances) of all four
(ferromagnetic, left-chiral, right-chiral, chiral spin-glass)
ordered phases, especially in the multicritical region. The chaotic
behavior of the interactions, under scale change, are determined in
the chiral spin-glass phase and on the boundary between the chiral
spin-glass and disordered phases, showing Lyapunov exponents in
magnitudes reversed from the usual ferromagnetic-antiferromagnetic
spin-glass systems.  At low temperatures, the boundaries of the
left- and right-chiral phases become thresholded in $p$ and $c$.  In
$d=2$, the chiral spin-glass Potts system does not have a spin-glass
phase, consistently with the lower-critical dimension of
ferromagnetic-antiferromagnetic spin glasses.  The left- and
right-chirally ordered phases show reentrance in chirality
concentration $p$.
    \end{abstract}
    \maketitle

\begin{figure}[ht!]
\centering
\includegraphics[scale=1.0]{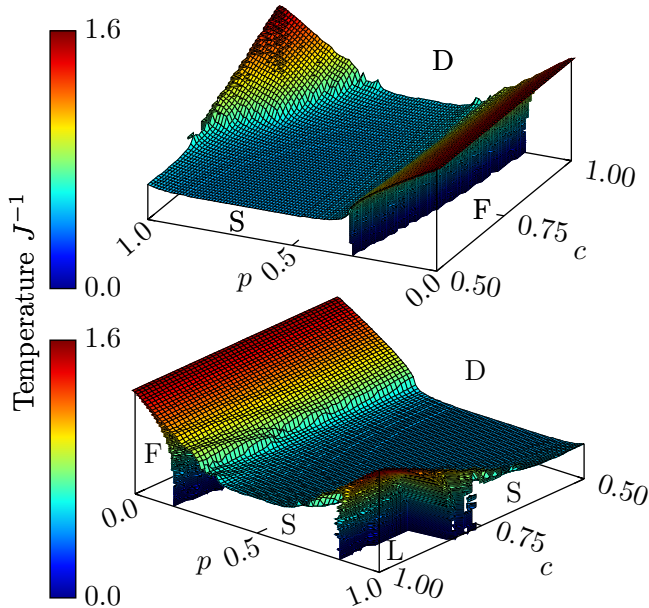}
\caption{(Color online) Calculated global phase diagram of the $d=3$
chiral Potts spin glass, in temperature $J^{-1}$, chirality
concentration $p$, and chirality-breaking concentration $c$. Note
that the upper and lower figures are rotated with respect to each
other. The ferromagnetically ordered phase (F), the chiral
spin-glass phase (S), the left-chirally ordered phase (L), and the
disordered phase (D) are marked.  The global phase diagram is
mirror-symmetric with respect to the chirality-breaking
concentration $c=0.5$, so that only $1\leq c \leq 0.5$ is shown.  In
the (not shown) mirror-symmetric $0.5\leq c \leq 0$ portion of the
global phase diagram, the right-chirally ordered phase (R) occurs in
the place of the left-chirally ordered phase (L) seen in this
figure. Different cross-sections of this global phase diagram are
shown in Figs. 2 and 3.}
\end{figure}

    \section{Introduction}
The chiral Potts model was originally introduced
\cite{Ostlund,Kardar,Huse,Huse2,Caflisch} to model the full phase
diagram of krypton monolayers, including the epitaxial and
incommensurate ordered phases. In addition to being useful in the
analysis of surface layers, the chiral Potts model has become an
important model of phase transitions and critical phenomena. We have
studied the chiral spin-glass Potts system with $q=3$ states in
$d=2$ and 3 spatial dimensions by renormalization-group theory and
calculated the global phase diagrams (Fig. 1) in temperature,
chirality concentration $p$, and chirality-breaking concentration
$c$, also quantitatively determining phase chaos and phase-boundary
chaos. In $d = 3$, the system has ferromagnetic, left-chiral,
right-chiral, chiral spin-glass, and disordered phases.  The phase
boundaries to the ferromagnetic, left- and right-chiral phases show,
differently, an unusual, fibrous patchwork (microreentrances) of all
four (ferromagnetic, left-chiral, right-chiral, chiral spin-glass)
ordered phases, especially in the multicritical region. The chaotic
behavior of the interactions, under scale change, is determined in
the chiral spin-glass phase and on the boundary between the chiral
spin-glass and disordered phases, showing Lyapunov exponents in
magnitudes reversed from the usual ferromagnetic-antiferromagnetic
spin-glass systems.  At low temperatures, the boundaries of the
left- and right-chiral phases become thresholded in $p$ and $c$.  In
the $d=2$, the chiral spin-glass Potts system does not have a
spin-glass phase, consistently with the lower-critical dimension of
ferromagnetic-antiferromagnetic spin glasses.  The left- and
right-chirally ordered phases show reentrance in chirality
concentration $p$.

\section{The Chiral Potts Spin-Glass System}
The chiral Potts model is defined by the Hamiltonian
\begin{equation}
\centering
- \beta {\cal H} = \sum_{\left<ij\right>}[J_0\delta(s_i,s_j) + J_\pm\delta(s_i, s_j\pm1)],
\end{equation}
where $\beta=1/k_{B}T$, at site $i$ the spin $s_{i}=1,2,,...,q$ can
be in $q$ different states with implicit periodic labeling, e.g.
$s_i=q+n$ implying $s_i=n$, the delta function $\delta(s_i,
s_j)=1(0)$ for $s_i=s_j (s_i\neq s_j)$, and $\langle ij \rangle$
denotes summation over all nearest-neighbor pairs of sites. The
upper and lower subscripts of $J_\pm>0$ give left-handed and
right-handed chirality (corresponding to heavy and superheavy domain
walls in the krypton-on-graphite incommensurate ordering
\cite{Kardar, Caflisch}), whereas $J_\pm=0$ gives the non-chiral
Potts model (relevant to the krypton-on-graphite epitaxial ordering
\cite{BerkerPLG}).

In the chiral Potts spin-glass model studied here, the chirality of
each nearest-neighbor interaction is randomly either left-handed, or
right-handed, or zero. This randomness is frozen (quenched) into the
system and the overall fraction of left-, right-, and non-chirality
is controlled by the quenched densities $p$ and $c$ as described
below. Thus, the Hamiltonian of the chiral Potts spin-glass model is
\begin{multline}
- \beta {\cal H} = \sum_{\left<ij\right>}J \,[(1-\eta_{ij})\delta(s_i, s_j)\\
+\eta_{ij}\,[\phi_{ij}\delta(s_i,s_j+1)+(1-\phi_{ij})\delta(s_i,s_j-1)],
\end{multline}
where, for each pair of nearest-neighbor sites $<ij>,$ $\eta_{ij}=0$
(non-chiral) or 1 (chiral).  In the latter case, $\phi_{ij}=1$
(left-handed) or 0 (right-handed).  Thus, non-chiral, left-chiral,
and right-chiral nearest-neighbor interactions are frozen randomly
distributed in the entire system.  For the entire system, the
overall concentration of chiral interactions is given by $p$, with
$0\leq p\leq1$.  Among the chiral interactions, the overall
concentrations of left- and right-chiral interactions are
respectively given by $c$ and $1-c$, with $0\leq c\leq1$. Thus, the
model is chiral for $p>0$ and chiral-symmetric $c=0.5$,
chiral-symmetry broken for $c\neq 0.5$. The global phase diagram is
given in terms of temperature $J^{-1}$, chirality concentration $p$,
and chirality-breaking concentration $c$.(Figs. 1-3)

Under the renormalization-group transformations described below, the
Hamiltonian given in Eq.(2) maps onto the more general form
\begin{multline}
- \beta {\cal H} = \sum_{\left<ij\right>} \, [J_0(ij)\delta(s_i,
s_j)+J_+(ij)\delta(s_i,s_j+1)\\
 +J_-(ij)\delta(s_i,s_j-1)],
\end{multline}
where for each pair of nearest-neighbor sites $<ij>$, the largest of
the interaction constants $(J_0,J_+,J_-)$ is set to zero, by
subtracting a constant G from each of $(J_0,J_+,J_-)$, with no
effect to the physics.

\section{Renormalization-Group Transformation: Migdal-Kadanoff Approximation / Exact Hierarchical Lattice Solution}

We solve the chiral Potts spin-glass model with $q=3$ states by
renormalization-group theory, in $d=3$ spatial dimension and with
the length rescaling factor $b=2$. Our solution is, simultaneously,
the Migdal-Kadanoff approximation \cite{Migdal,Kadanoff} for the
cubic lattices and exact
\cite{BerkerOstlund,Kaufman1,Kaufman2,McKay,Hinczewski1} for the
$d=3$ hierarchical lattice based on the leftmost graph of Fig. 4.
Exact calculations on hierarchical lattices
\cite{BerkerOstlund,Kaufman1,Kaufman2,McKay,Hinczewski1} are also
currently widely used on a variety of statistical mechanics problems
\cite{Kaufman,Kotorowicz,Barre,Monthus,Zhang,Shrock,Xu,Hwang2013,Herrmann1,
Herrmann2,Garel,Hartmann,Fortin,Wu,Timonin,Derrida,Thorpe,Efrat,Monthus2,
Hasegawa,Lyra,Singh,Xu2014,Hirose1,Silva,Hotta,Boettcher1,Boettcher2,Hirose2,Boettcher3,Nandy}.
This approximation for the cubic lattice is an uncontrolled
approximation, as in fact are all renormalization-group theory
calculations in $d=3$ and all mean-field theory calculations.
However, as noted before \cite{Yunus}, the local summation in
position-space technique used here has been qualitatively,
near-quantitatively, and predictively successful in a large variety
of problems, such as arbitrary spin-$s$ Ising models
\cite{BerkerSpinS}, global Blume-Emery-Griffiths model
\cite{BerkerWortis}, first- and second-order Potts transitions
\cite{NienhuisPotts,AndelmanBerker}, antiferromagnetic Potts
critical phases \cite{BerkerKadanoff1,BerkerKadanoff2}, ordering
\cite{BerkerPLG} and superfluidity \cite{BerkerNelson} on surfaces,
multiply reentrant liquid crystal phases \cite{Indekeu,Garland},
chaotic spin glasses \cite{McKayChaos}, random-field
\cite{Machta,FalicovRField} and random-temperature
\cite{HuiBerker,HuiBerkerE} magnets including the remarkably small
$d=3$ magnetization critical exponent $\beta$ of the random-field
Ising model, and high-temperature superconductors
\cite{HincewskiSuperc}. Thus, this renormalization-group
approximation continues to be widely used
\cite{Gingras1,Migliorini,Gingras2,Hinczewski,Heisenberg,Guven,Ohzeki,Ozcelik,Gulpinar,Kaplan,Ilker1,Ilker2,Ilker3,Demirtas}.

\begin{figure*}[ht!]
\centering
\includegraphics[scale=1.0]{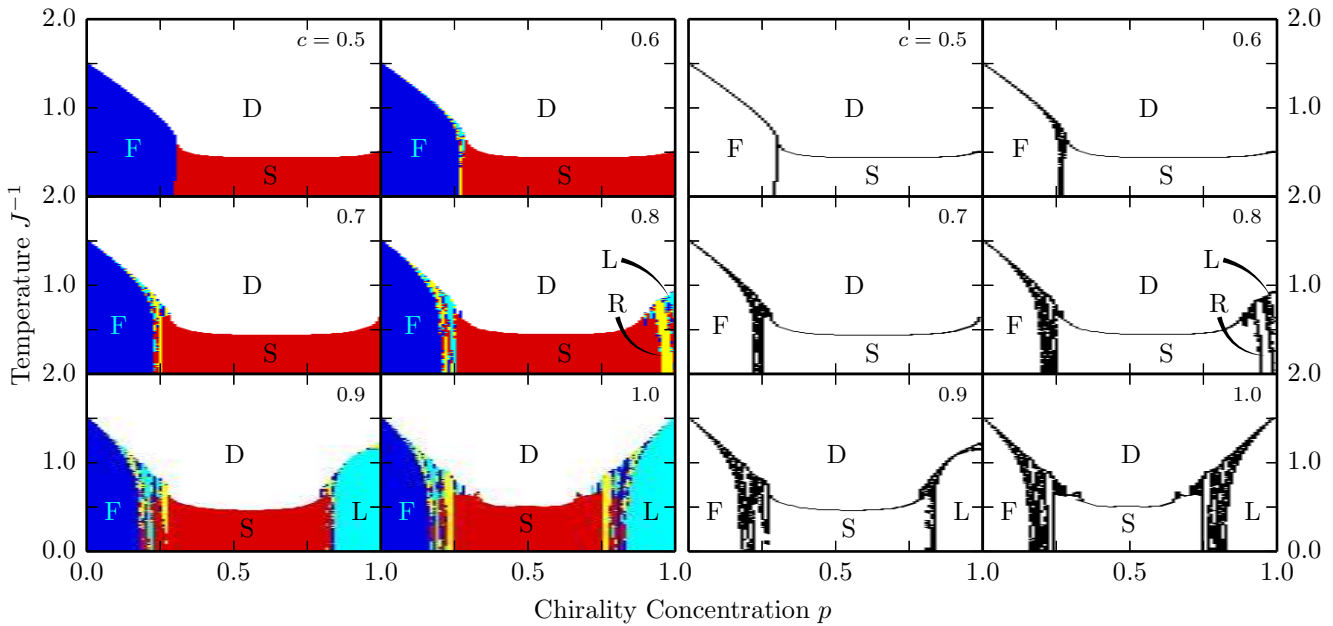}
\caption{(Color online) Cross-sections, in temperature $J^{-1}$ and
chirality concentration $p$, of the global phase diagram shown in
Fig. 1. The chirality-breaking concentration $c$ is given on each
cross-section. The ferromagnetically ordered phase (F), the chiral
spin-glass phase (S), the left- and right-chirally ordered phases (L
and R), and the disordered phase (D) are marked.  Note that, as soon
as the chiral symmetry of the model is broken by $c \neq 0.5$, a
narrow fibrous patchwork (microreentrances) of all four
(ferromagnetic, left-chiral, right-chiral, chiral spin-glass)
ordered phases intervenes at boundaries of the ferromagnetically
ordered phase F. This intervening region is more pronounced close to
the multicritical region where the ferromagnetic, spin-glass, and
disordered phases meet. The interlacing phase transitions inside
this region are more clearly seen in the right-hand side panels of
the figure, where only the phase boundaries are drawn in black. This
intervening region gains importance as $c$ moves away from 0.5. But
it is only at higher values of the chirality-breaking concentration
$c$, such as $c=0.8$ on the figure, that the chirally ordered phase
appears as a compact region at $c,p\lesssim 1$. In this case, again
all four (ferromagnetic, left-chiral, right-chiral, chiral
spin-glass) ordered phases intervene in a narrow fibrous patchwork
at the boundaries of the chirally ordered phase L and R, the latter
mirror-symmetric and not shown here. For $c=1$, for which all
interactions of the system are, with respective concentrations $1-p$
and $p$, either ferromagnetic, or left-chiral, the phase diagram
becomes symmetric with respect to $p=0.5$ as in standard
ferromagnetic-antiferromagnetic spin-glass systems, except that the
chirally ordered phases dominate the fibrous patchwork on both sides
of the phase diagram.}
\end{figure*}

\begin{figure*}[ht!]
\centering
\includegraphics[scale=1.02]{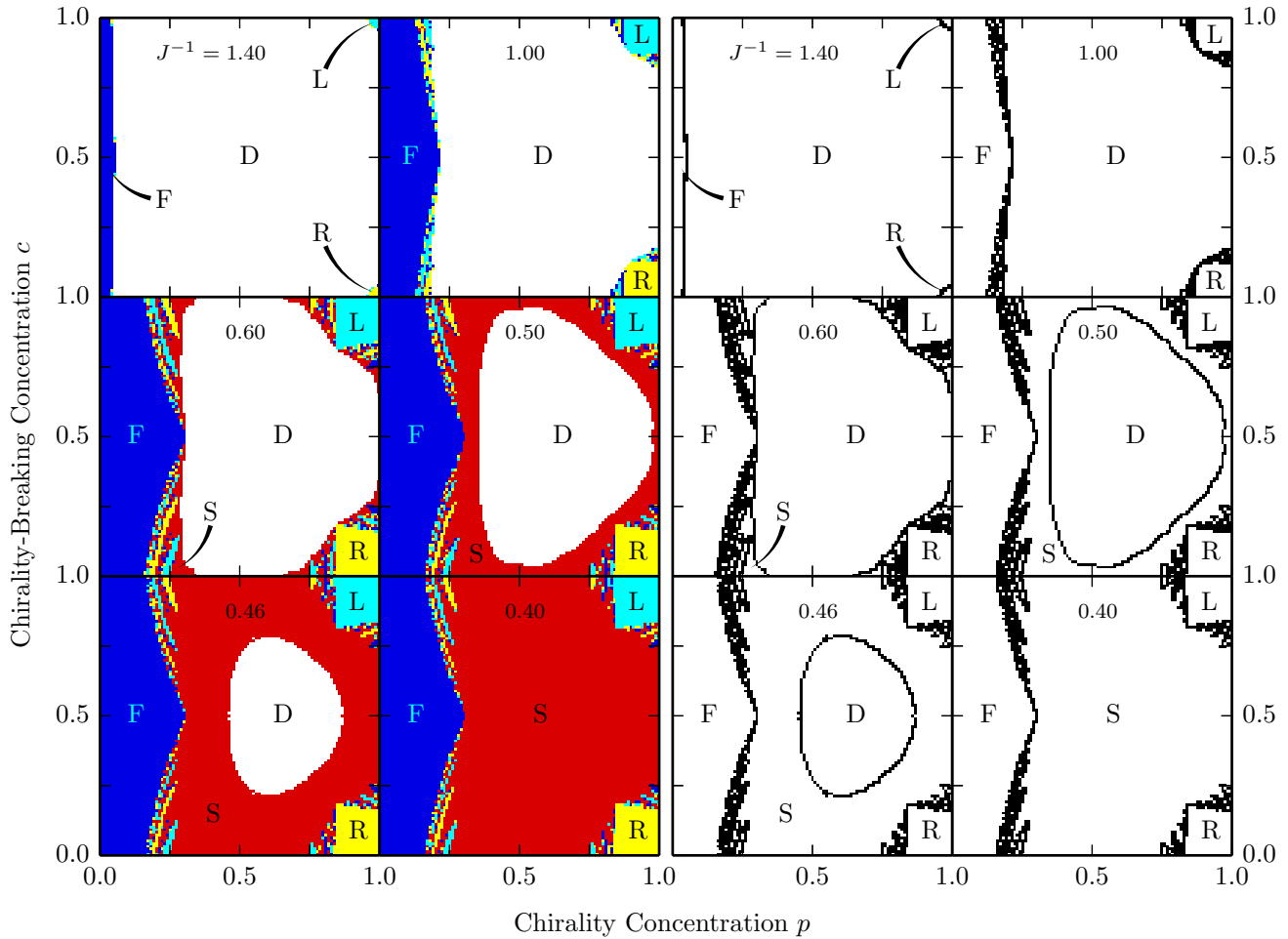}
\caption{(Color online) Cross-sections, in chirality concentration
$p$ and chirality-breaking concentration $c$, of the global phase
diagram shown in Fig. 1. The temperature $J^{-1}$ is given on each
cross-section. The ferromagnetically ordered phase (F), the chiral
spin-glass phase (S), the left-chirally ordered phase (L), the
right-chirally ordered phase (R), and the disordered phase (D) are
marked. Note the narrow fibrous patches (microreentrances) of all
four (ferromagnetic, left-chiral, right-chiral, chiral spin-glass)
ordered phases intervening at the boundaries of the
ferromagnetically ordered phase F and at the boundaries of the
chirally ordered phases L and R. It is seen here that, within these
regions, the chirally ordered phases L and R form elongated lamellar
patterns. These intervening phase transitions are more clearly seen
in the right-hand side panels of the figure, where only the phase
boundaries are drawn in black. Also note the temperature-independent
square shape, at low temperatures, of the phase boundary of the
chirally ordered phases, creating thresholds of $p = 0.84$ and $c =
0.84$ or 0.16 into L or R, respectively. This is also visible in the
three-dimensional Fig. 1}
\end{figure*}

\begin{figure}[ht!]
\centering
\includegraphics[scale=1.0]{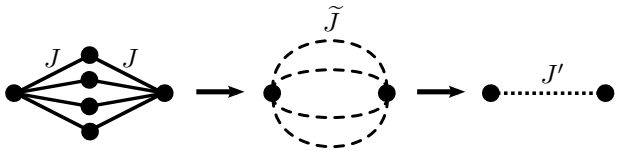}
\caption{Renormalization-group transformation consisting of
decimation followed by bond moving.  The resulting recursion
relations are approximate for the cubic lattice.  The corresponding
hierarchical lattice is obtained by the repeated self-imbedding of
the leftmost graph.  The recursion relations are exact for this $d =
3$ hierarchical lattice. For the $d = 2$, the number of parallel
strands is 2 instead of 4 shown here.}
\end{figure}

The local renormalization-group transformation is achieved by a
sequence, shown in Fig. 4, of decimations
\begin{multline}
\begin{split}
e^{\widetilde{J}_0(13)-\widetilde{G}}= x_0(12)x_0(23) &+ x_+(12)x_-(23)\\
 &+ x_-(12)x_+(23),\\
e^{\widetilde{J}_+(13)-\widetilde{G}}= x_0(12)x_+(23) &+ x_+(12)x_0(23)\\
&+ x_-(12)x_-(23),\\
e^{\widetilde{J}_-(13)-\widetilde{G}}= x_0(12)x_-(23) &+ x_-(12)x_0(23)\\
&+x_+(12)x_+(23),\\
\end{split}
\end{multline}
where $x_0(12) \equiv e^{J_0(12)}$, etc., and $\widetilde{G}$ is the
subtractive constant mentioned in the previous section, and bond
movings
\begin{multline}
\begin {split}
J_0'(13) =&
\widetilde{J}_0^{(1)}(13)+\widetilde{J}_0^{(2)}(13)+\widetilde{J}_0^{(3)}(13)+\widetilde{J}_0^{(4)}(13),\\
J_+'(13) =&
\widetilde{J}_+^{(1)}(13)+\widetilde{J}_+^{(2)}(13)+\widetilde{J}_+^{(3)}(13)+\widetilde{J}_+^{(4)}(13),\\
J_-'(13) =&
\widetilde{J}_-^{(1)}(13)+\widetilde{J}_-^{(2)}(13)+\widetilde{J}_-^{(3)}(13)+\widetilde{J}_-^{(4)}(13),
\end {split}
\end{multline}
where primes mark the interactions of the renormalized system.

\begin{figure}[ht!]
\centering
\includegraphics[scale=1.0]{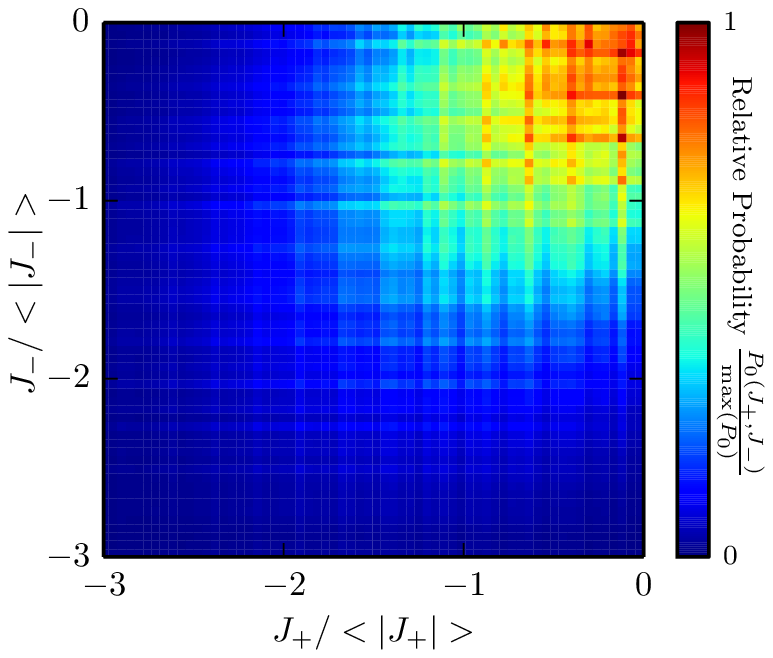}
\caption{(Color online) The fixed probability distribution of the
quenched random interactions $P_0(J_+,J_-)$ to which all of the
points in the chiral spin-glass phase are attracted under
renormalization-group transformations, namely the sink of the chiral
spin-glass phase.  The average interactions $<J_\pm>$ diverge to
negative infinity as $<J_\pm> \sim b^{y_R n}$, where $n$ is the
number of renormalization-group iterations and $y_R = 0.32$ is the
runaway exponent, while $J_0=0$ (See Sec. II). The other two
distributions $P_+(J_0,J_-)$ and $P_-(J_0,J_+)$ have the same sink
distribution. Thus, in the chiral spin-glass phase, chiral symmetry
is broken by local order, but not globally.}
\end{figure}

The starting trimodal quenched probability distribution of the
interactions, characterized by $p$ and $c$ as described above, is
not conserved under rescaling. The renormalized quenched probability
distribution of the interactions is obtained by the convolution
\cite{Andelman}
\begin{multline}
P'(\textbf{J}(i'j')) = \\
\int{\left[\prod_{ij}^{i'j'}d\textbf{J}(ij)P(\textbf{J}(ij))\right]}
\delta(\textbf{J}(i'j')-\textbf{R}(\left\{\textbf{J}(ij)\right\})),
\end{multline}
where $\textbf{J}\equiv (J_0',J_+',J_-')$ and
$\textbf{R}(\left\{\textbf{J}(ij)\right\})$ represents the bond
decimation and bond moving given in Eqs.(4) and (5).  Similar
previous studies, on other spin-glass systems, are in Refs.
\cite{Gingras1,Migliorini,Gingras2,Hinczewski,Heisenberg,Guven,Ohzeki,Ozcelik,Gulpinar,Kaplan,Ilker1,Ilker2,Ilker3,Demirtas}.

For numerical practicality, the bond moving of Eq. (5) is achieved
by two sequential pairwise combination of interactions, each
pairwise combination leading to an intermediate probability
distribution resulting from a pairwise convolution as in Eq.(6).
Furthermore, due to our convention of zeroing the largest
interaction constant in each local triplet of interactions, the
quenched probability distribution of three interactions
$P(\textbf{J}(ij))$ is conveniently just composed of the three
probability distributions of two interactions,
$P_0(J_+,J_-),P_+(J_0,J_-),P_-(J_+,J_-)$, where $P_0(J_+,J_-)$ has
the (largest) interaction $J_0 =0,$ etc., which also considerably
simplifies the numerical calculation. We effect this procedure
numerically, by representing each probability distribution by
histograms, as in previous studies
\cite{Migliorini,Hinczewski,Heisenberg,Guven,Ozcelik,Gulpinar,Ilker2,Demirtas}.
The probability distributions of two interactions $P_0(J_+,J_-)$,
$P_+(J_0,J_-)$, and $P_-(J_+,J_-)$ are represented via bivariate
histograms with two-dimensional vectors $(J_+,J_-)$ for $P_0$, etc.
The number of histograms grow rapidly with each
renormalization-group transformation, so that for calculational
purposes, the histograms are binned when the number of histograms
outgrow $40,000$ bins. In the calculation of chiral spin-glass
phase-sink fixed distribution of Fig. 5, the histograms are binned
after $10^8$ histograms.

The different thermodynamic phases of the model are identified by
the different asymptotic renormalization-group flows of the quenched
probability distribution.  For all renormalization-group flows,
originating inside the phases and on the phase boundaries, Eq.(6) is
iterated until asymptotic behavior is reached. Thus, we are able to
calculate the global phase diagram of the chiral Potts spin-glass
model.

\section{Chiral Potts Spin Glass: Calculated Global Phase Diagram}

The calculated global phase diagram of the $d=3$ chiral Potts
spin-glass system, in temperature $J^{-1}$, chirality concentration
$p$, and chirality-breaking concentration $c$, is given in Fig. 1.
The ferromagnetically ordered (F) phase occurs at low temperature
and low chirality $p$. The chiral spin-glass ordered (S) phase
occurs at intermediate chirality $p$ for all $c$ and at high
chirality $p$ for intermediate $c$. The left- and right-chirally
ordered phases L and R occur at high chirality $p$ and values of
chirality-breaking $c$ away from 0.5. The disordered phase (D)
occurs at high temperature. The global phase diagram is
mirror-symmetric with respect to the chirality-breaking
concentration $c=0.5$, so that only $1\leq c \leq 0.5$ is shown in
Fig. 1. In the (not shown) mirror-symmetric $0.5\leq c \leq 0$
portion of the global phase diagram, the right-chirally ordered
phase (R) occurs in the place of the left-chirally ordered phase (L)
seen in Fig. 1. Different cross-sections of the global phase diagram
are shown in Figs. 2 and 3.

Under renormalization-group transformations, all points in the
spin-glass phase are attracted to a fixed probability distribution
of the quenched random interactions $P(J_0,J_+,J_-)$, namely to the
sink of the chiral spin-glass phase. As explained in Sec. III,
$P(J_0,J_+,J_-)$ is composed of three distributions, $P_0(J_+,J_-)$,
$P_+(J_0,J_-)$, and $P_-(J_0,J_+)$. Of these, $P_0(J_+,J_-)$ gives
the quenched probability distribution of nearest-neighbor
interactions in which the ferromagnetic interaction $J_0$ is
dominant. Similarly, $P_+(J_0,J_-)$ and $P_-(J_0,J_+)$ give the
quenched probability distributions of nearest-neighbor interactions
in which, respectively, the left-chiral interaction $J_+$ and the
right-chiral interaction $J_-$ are dominant. (As explained in Sec.
II, by subtraction of an overall constant, the dominant interaction
is set to zero and the other two, subdominant interactions are
therefore negative, with no loss of generality.) The sink fixed
distribution for $P_0(J_+,J_-)$ is given in Fig. 5, where the
average interactions $<J_\pm>$ diverge to negative infinity as
$b^{y_R n}$, where $n$ is the number of renormalization-group
iterations and $y_R = 0.32$ is the runaway exponent, while
conserving the shape of the distribution shown in Fig. 5. The other
two distribution $P_+(J_0,J_-)$ and $P_-(J_0,J_+)$ have the same
sink distribution. Thus, in the chiral spin-glass phase, chiral
symmetry is broken by local order, but not globally.

In spin-glass phases, at a specific location in the lattice, the
consecutive interactions, encountered under consecutive
renormalization-group transformations, behave chaotically
\cite{McKayChaos, McKayChaos2,BerkerMcKay}. This chaotic behavior
was found \cite{McKayChaos, McKayChaos2,BerkerMcKay} and
subsequently well established \cite {Bray,Hartford,
Nifle1,Nifle2,Banavar,Frzakala1,Frzakala2,Sasaki,Lukic,Ledoussal,Rizzo,
Katzgraber,Yoshino,Pixley,Aspelmeier1,Aspelmeier2,Mora,Aral,Chen,Jorg,Lima,Katzgraber2,MMayor,ZZhu,Katzgraber3,Fernandez,Ilker1,Ilker2}
in spin-glass systems with competing ferromagnetic and
antiferromagnetic interactions.  We find here that the chaotic
rescaling behavior also occurs in our current spin-glass system with
competing left- and right-chiral interactions, as shown in Fig. 6.
In fact, the chaotic rescaling behavior occurs not only within the
spin-glass phase, but also, quantitatively distinctly, at the phase
boundary between the spin-glass and disordered phases \cite{Ilker1}.
This chaotic behavior at the phase boundary is also seen in the
chiral system here and also shown in Fig. 6. It has been shown that
chaos in the interaction as a function of rescaling implies chaos in
the spin-spin correlation function as a function of distance
\cite{Aral}. Chaos in the spin-glass phase and at its phase boundary
are identified and distinguished by different Lyapunov exponents
\cite{Aral,Ilker1,Ilker2}.  We have calculated the Lyapunov exponent
\cite{Collet, Hilborn}
\begin{equation}
\lambda = \lim _{n\rightarrow\infty} \frac{1}{n} \sum_{k=0}^{n-1}
\ln \Big|\frac {dx_{k+1}}{dx_k}\Big|
\end{equation}
where $x_k = J(ij)/<J>$ at step $k$ of the renormalization-group
trajectory.  The sum in Eq.(7) is to be taken within the asymptotic
chaotic band, which is renormalization-group stable or unstable for
the phase or its boundary, respectively. Thus, we throw out the
first 100 renormalization-group iterations to eliminate the
transient points outside of, but leading to the chaotic band.
Subsequently, typically using 1,000 renormalization-group iterations
in the sum in Eq.(7) assures the convergence of the Lyapunov
exponent value.  Thus, the Lyapunov exponents that we obtain are
numerically exact, to the number of digits given. We have calculated
the Lyapunov exponents $\lambda = 1.77$ and 1.94 respectively for
the chiral spin-glass phase and for the boundary between the chiral
spin-glass and disordered phases. At the chiral spin-glass
phase-sink fixed distribution, the average interaction diverges to
negative infinity as $<J> \sim b^{ny_R}$, where $n$ is the number of
renormalization-group iterations and $y_R = 0.32$ is the runaway
exponent. At the fixed distribution of the phase boundary between
the chiral spin-glass and disordered phases, the average interaction
remains fixed at $<J> = -2.53$. Interestingly, chaos is stronger at
the boundary (larger Lyapunov exponent) than inside the chiral
spin-glass phase. The opposite is seen in the usually studied $\pm
J$ ferromagnetic-antiferromagnetic spin glass \cite{Ilker1}.

\begin{figure}[ht!]
\centering
\includegraphics[scale=1.0]{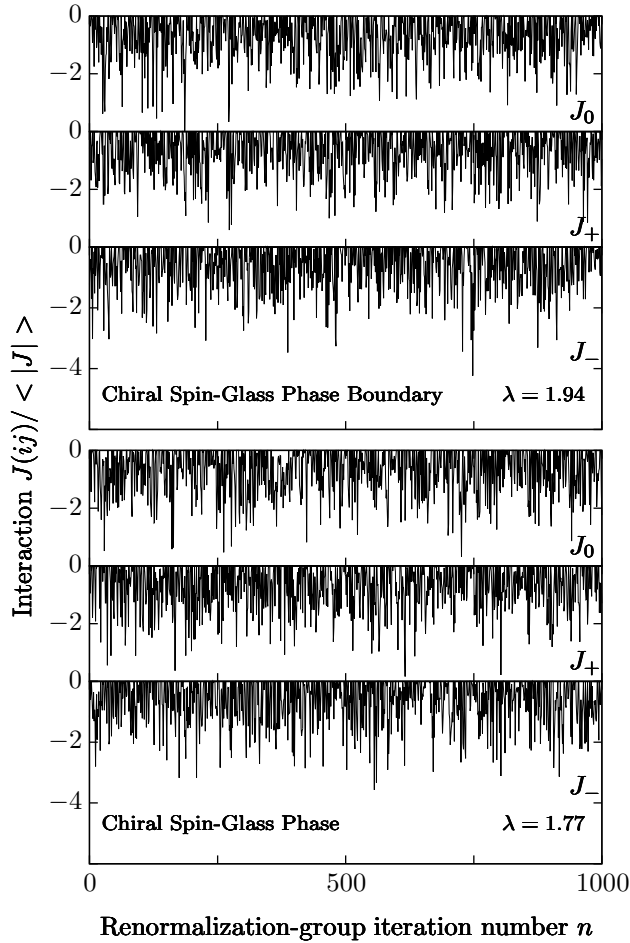}
\caption{Chaotic renormalization-group trajectory: The three
interactions at a given location, under consecutive
renormalization-group transformations, are shown. Bottom panel:
Inside the chiral spin-glass phase. The corresponding Lyapunov
exponent is $\lambda = 1.77$ and the average interaction diverges as
$<J> \sim b^{y_R n}$, where $n$ is the number of
renormalization-group iterations and $y_R = 0.32$ is the runaway
exponent. Top panel: At the phase boundary between the chiral
spin-glass and disordered phases. The corresponding Lyapunov
exponent is $\lambda = 1.94$ and the average non-zero interaction is
fixed at $<J> = -2.53$. The relative value of the Lyapunov exponents
is unusual for spin-glass systems.}
\end{figure}

By contrast, in each of the ferromagnetic (F), left-chiral (L), and
right-chiral (R) ordered phases, under consecutive
renormalization-group transformations, the quenched probability
distribution of the interactions sharpens to a delta function around
a single value receding to negative infinity, for the respective
pairs of interactions, namely ($J_+,J_-), (J_0, J_+)$, and $(J_0,
J_-)$. There is no asymptotic chaotic behavior under
renormalization-group in these phases F, L, and R.

\begin{figure*}[ht!]
\centering
\includegraphics[scale=1.0]{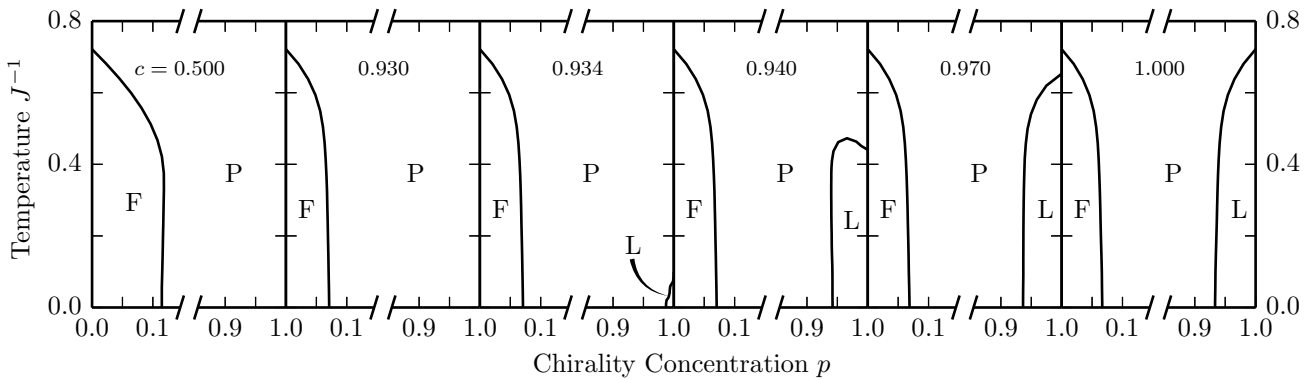}
\caption{Representative cross-sections of the $d=2$ chiral Potts
spin-glass system, in temperature $J^{-1}$ and chirality
concentration $p$. The chirality-breaking concentration $c$ is given
on each cross-section. The ferromagnetically ordered phase (F), the
left-chirally ordered phase (L), and the disordered phase (D) are
marked. No chiral spin-glass phase occurs in $d=2$ and no fibrous
patchwork is seen at the phase boundaries. The chirally ordered
phase appears for very high chirality-breaking concentration $c$
(seen here for $c = 0.934$, but not seen for $c = 0.930$) and shows
reentrance in chirality concentration $p$. This reentrance
disappears as $c = 1$ is approached. For $c=1$, for which all
interactions of the system are, with respective concentrations $1-p$
and $p$, either ferromagnetic, or left-chiral, the phase diagram
becomes symmetric with respect to $p=0.5$ as in standard
ferromagnetic-antiferromagnetic spin-glass systems.}
\end{figure*}

Cross-sections of the global phase diagram, in temperature $J^{-1}$
and chirality concentration $p$, are given in Fig. 2. The
chirality-breaking concentration $c$ is indicated for each
cross-section. Note that, as soon as the chiral symmetry of the
model is broken by $c \neq 0.5$, a narrow fibrous patchwork
(microreentrances) of all four (ferromagnetic, left-chiral,
right-chiral, chiral spin-glass) ordered phases intervenes at the
boundaries between the ferromagnetically ordered phase F and the
spin-glass phase S or the disordered phase D. This intervening
region is more pronounced close to the multicritical region where
the ferromagnetic, spin-glass, and disordered phases meet. The
interlacing phase transitions inside this region are more clearly
seen in the right-hand side panels of Fig. 2, where only the phase
boundaries are drawn in black. This intervening region gains
importance as $c$ moves away from 0.5. But it is only at higher
values of the chirality-breaking concentration $c$, such as $c=0.8$
on the figure, that the chirally ordered phase appears as a compact
region at $c,p\lesssim 1$. In this case, again all four
(ferromagnetic, left-chiral, right-chiral, chiral spin-glass)
ordered phases intervene in a narrow fibrous patchwork at the
boundaries of the chirally ordered phases L and R, the latter mirror
symmetric and not shown here. For $c=1$, for which all interactions
of the system are, with respective concentrations $1-p$ and $p$,
either ferromagnetic, or left-chiral, the phase diagram becomes
symmetric with respect to $p=0.5$ as in standard
ferromagnetic-antiferromagnetic spin-glass systems
\cite{NishimoriBook}, except that the chirally ordered phases
dominate the fibrous patchwork on both sides of the phase diagram.

Cross-sections, in chirality concentration $p$ and
chirality-breaking concentration $c$, of the global phase diagram
are given in Fig. 3. The temperature $J^{-1}$ is given on each
cross-section. Note the narrow fibrous patches of all four
(ferromagnetic, left-chiral, right-chiral, chiral spin-glass) phases
intervening at the boundaries of the ferromagnetically ordered phase
F and at the boundaries of the chirally ordered phases L and R. It
is seen here that, within these regions, the chirally ordered phases
L and R form elongated lamellar patterns. The interlacing phase
transitions inside this region are more clearly seen in the
right-hand side panels of the figure, where only the phase
boundaries are drawn in black. It is again seen that the symmetry
around $p=0.5$ at the upper horizontal frame $(c=1)$ of each panel
is broken inside the panel $(c<1)$. Also note the
temperature-independent square shape, at low temperatures, of the
phase boundary of the chirally ordered phases L and R, creating the
threshold value of $p = 0.84$ and $c = 0.84$ or 0.16 into L or R,
respectively.  This is also visible in the three-dimensional Fig. 1.

\section{Chiral Reentrance in $d=2$}

The global phase diagram of the $d=2$ chiral Potts spin-glass system
is given in Fig. 7. Representative cross-sections in temperature
$J^{-1}$ and chirality concentration $p$ are shown. The
chirality-breaking concentration $c$ is given on each cross-section.
The ferromagnetically ordered phase (F), the left-chirally ordered
phase (L), and the disordered phase (D) are marked. No chiral
spin-glass phase occurs in $d=2$ and no fibrous patchwork is seen at
the phase boundaries. The chirally ordered phase appears for very
high chirality-breaking concentration $c$ (seen here for $c =
0.934$, but not seen for $c = 0.930$) and shows reentrance
\cite{Cladis, Hardouin, Indekeu, Garland, Netz, Kumari, Caflisch} in
chirality concentration $p$. This reentrance disappears as $c = 1$
is approached. For $c=1$, for which all interactions of the system
are, with respective concentrations $1-p$ and $p$, either
ferromagnetic, or left-chiral, the phase diagram becomes symmetric
with respect to $p=0.5$ as in standard
ferromagnetic-antiferromagnetic spin-glass systems \cite{Ilker2}.

The absence of the chiral spin-glass phase in $d = 2$ is consistent
with standard ferromagnetic-antiferromagnetic Ising spin-glass
systems, where the lower-critical dimension for the spin-glass phase
is found around 2.5
\cite{Parisi,Boettcher,Amoruso,Bouchaud,Demirtas}. Below this
dimension, no spin-glass phase appears (unless some
nano-restructuring is done to the system \cite{Ilker2}).

\section{Conclusion}

We have thus obtained the global phase diagram of the chiral
spin-glass Potts system with $q=3$ states in $d=3$ and 2 spatial
dimensions by renormalization-group theory that is approximate for
the cubic lattice and exact for the hierarchical lattice.  Unusual
features have been revealed in $d=3$. The phase boundaries to the
ferromagnetic, left- and right-chiral phases show, differently, an
unusual, fibrous patchwork (microreentrances) of all four
(ferromagnetic, left-chiral, right-chiral, chiral spin-glass)
ordered phases, especially in the multicritical region. In $d=3$,
there is a chiral spin-glass phase. Quite unusually, the phase
boundary between the chiral spin-glass and disordered phases is more
chaotic than the chiral spin-glass phase itself, as judged by the
magnitudes of the respective Lyapunov exponents. At low
temperatures, the boundaries of the left- and right-chiral phases
become temperature-independent and thresholded in chirality
concentration $p$ and chirality-breaking concentration $c$. In the
$d=2$, the chiral spin-glass system does not have a spin-glass
phase, consistently with the lower-critical dimension of
ferromagnetic-antiferromagnetic spin glasses.  The left- and
right-chirally ordered phases show reentrance in chirality
concentration $p$.

\begin{acknowledgments}
Support by the Academy of Sciences of Turkey (T\"UBA) is gratefully
acknowledged.
\end{acknowledgments}

\end{document}